\documentclass[%
 reprint,
longbibliography,
 amsmath,amssymb,
 aps,
prl,
floatfix,
]{revtex4-2}

\usepackage{graphicx}% Include figure files
\usepackage{dcolumn}% Align table columns on decimal point
\usepackage{bm}% bold math
\usepackage{braket}
\usepackage{color}

\begin{document}

%\preprint{APS/123-QED}

\title{Thermal Ising transition in two-dimensional SU(3) Fermi lattice gases with population imbalance
}% Force line breaks with \\

\author{Hayato Motegi,$^{\rm{1}}$ Giacomo Marmorini,$^{\rm{1,2}}$ Nobuo Furukawa,$^{\rm{1}}$ and Daisuke Yamamoto$^{\rm{2}}$}
\email[Corresponding author: ]{yamamoto.daisuke21@nihon-u.ac.jp}
\affiliation{%
$^1$Department of Physics and Mathematics, Aoyama Gakuin University, Sagamihara, Kanagawa 252-5258, Japan \\
$^2$Department of Physics, Nihon University, Tokyo 156-8550, Japan
}%

\begin{abstract}

We focus on three-component SU(3) Fermi gases loaded into a square optical lattice, with population imbalance between one component and the others. At strong coupling the system is described by the SU(3) Heisenberg model with an external field that couples to the population imbalance. We discuss the ground state at the mean-field level and then analyze the thermal fluctuations with the semi-classical Monte Carlo method. The interplay of interactions, population imbalance and thermal fluctuations gives rise to a phase transition linked to the breaking of an emergent Ising symmetry, despite the absence of frustration.
%The competition between antiferromagnetic interactions and the population imbalance gives rise to a thermal phase transition in the Ising universality class. 
This represents a new scenario of discrete symmetry breaking in low-dimensional systems with continuous symmetries. Possible implementations with cold alkaline-earth(-like) atoms are discussed.

%We investigate the magnetic ordering in three-component SU(3) Fermi gases loaded into a square optical lattice. We consider the case with population imbalance between one component and the others, for which we study the quantum and thermal phase transition phenomena in the presence of the population imbalance field. We discuss the low-temperature properties using the linear flavor-wave theory, which takes into account linear fluctuations from mean-field ordered states. In addition, we analyze the thermal phase transition phenomena with the semi-classical Monte-Carlo method. The competition between antiferromagnetic interactions and the population imbalance gives rise to several interesting magnetic states. Of particular interest among them is a true long-range ordered state, favored by the thermal fluctuations, which preserves the continuous symmetry of the Hamiltonian. 

%ちょっと付け足してみた
% experimental implications for the Mott-insulating states of three-flavor fermionic atoms in optical lattices are discussed.

\end{abstract}

\maketitle

%\tableofcontents
\textit{Introduction.}
Symmetry and its spontaneous breaking have been playing a central role for understanding the low-energy physics of many-body systems and  classifying phase transition phenomena. According to the Mermin-Wagner-Hohenberg theorem~\cite{Hohenberg, MerminWagner}, continuous symmetries cannot be spontaneously broken at nonzero temperature in one and two-dimensional systems with sufficiently short-range interactions.
%; the formation of true long-range order in a continuous order parameter is prohibited in such settings. 
However, it has been found in certain systems with only continuously symmetric interactions that a spontaneous symmetry breaking with respect to a discrete order parameter can emerge through a nontrivial mechanism even in low dimensions at nonzero temperature. In the framework of Heisenberg-like models, in a seminal work~\cite{chandra-90} Chandra, Coleman and Larkin introduced the idea that this scenario can be realized in the presence of frustration coming from competing exchange interaction; in fact, their proposal of an Ising transition in the frustrated $J_1$-$J_2$ model on the square lattice has been confirmed in various subsequent studies~\cite{weber-03,capriotti-04,gauthe-22}. 
An additional example of this kind is given by the Ising transition in the fully frustrated spin-1/2 Heisenberg ferromagnetic/antiferromagnetic square bilayer,
which exhibits a finite temperature phase transition in the 2D Ising universality class
~\cite{Egergent_discrete_symmetry1,Egergent_discrete_symmetry2},  occurring at the endpoint of the discontinuous (first-order) phase transition between the singlet-dimer and fully polarized triplet phases. 
%%The liquid-gas transition is another classic example of such an emergent Ising criticality. 
In the triangular lattice Heisenberg model it is the interplay of geometric frustration and magnetic field that stabilizes the up-up-down state, 
%selected by both quantum and thermal fluctuations out of the classically degenerate ground-state manifold; this state 
which breaks only a discrete translational $Z_3$ symmetry. In this Letter we aim to extend these concepts to SU($\mathcal{N}$)-symmetric Heisenberg models: we argue that, even without frustration coming from geometry or competing exchange interactions, the presence of a suitable ``external field'' [the SU($\mathcal{N}$) symmetry admits $\mathcal{N}-1$ couplings that play the role of generalized magnetic fields] can indeed induce the breaking of an emergent discrete symmetry.  

%In a magnetic field, the isotropic Heisenberg model on the triangular lattice exhibits a large accidental classical ground-state degeneracy. Thermal or quantum fluctuations lift the extensive degeneracy in favor of the so-called ``up-up-down state'', which breaks only a discrete translational $Z_3$ symmetry. A complex interplay of exchange interactions with dimensionality, geometry, and fluctuations can thus give rise to spontaneous breaking of an emergent discrete symmetry in systems with continuously symmetric interactions.
%\par
{The optimal experimental platform to test the above idea is given by ultracold atoms in optical lattices, which represent invaluable quantum simulators of many-body physics.
%Due to the high  controllability of various parameters including interaction, dimensionality, and population, they have been established as invaluable quantum simulators of many-body physics.
%studying phase transition phenomena in quantum many-body systems. 
For instance,  two-component Fermi atoms in an optical lattice can realize the Hubbard model~\cite{Short_range_antiferro}, the simplest model for strongly correlated electrons. At half filling and strong coupling the system is well approximated by the SU(2) Heisenberg model~\cite{order_parameter_space}, and
%and, for bipartite lattices in two or more dimensions, the ground state is the conventional antiferromagnetic state.
%whose order parameter space is a two-dimensional sphere ${\rm S}^2$ stemming from the SU(2) spin rotational symmetry~\cite{order_parameter_space}. In fact, the 
recent experiments employing the quantum gas microscope technique have confirmed that in a square optical lattice short-range spin-spin correlations exhibit finite correlation length, as expected at low but finite temperatures ($\gtrsim$ 0.25 times the tunneling energy)~\cite{Short_range_antiferro}.} Owing to the  advances in the manipulation of cold alkaline-earth(-like) atoms, such as $ ^{173}{\rm Yb} $ and $ ^{87}{\rm Sr} $, analogous experiments on systems with SU($\mathcal{N}$) symmetry, $\mathcal{N}>2$, have been underway~\cite{hofrichter-16,taie-22,cazalilla-14,takahashi-22}. This has stimulated vast theoretical work, leading to numerous predictions of exotic ground states for various lattice geometries and and degrees $\mathcal{N}$~\cite{SU(3)_1,SU(3)_2,corboz-11,sotnikov-14,hafez-19}. {However, the effects of ``external fields'' that partially break the SU($\mathcal{N}$) symmetry have been rarely investigated, as well as those of thermal fluctuations~\cite{Yamamotosan}.}
%, in contrast to the SU(2) spin systems under  magnetic field. 
Note that in cold-atom experiments, such an external-field effect can be simulated by imposing a global imbalance of populations among the $\mathcal{N}$ components~\cite{spinimbalance}.

\par
%\textit{SU(3) lattice fermions with population imbalance.---}  
\textit{Model.}
Inspired by the previous considerations, in this Letter we study the 
%phase transition phenomena in the 
strong coupling regime of three-component Fermi gases with SU(3)-symmetric interactions in a square optical lattice at 1/3 filling~\cite{SU(3)_1,SU(3)_2} with population imbalance between one component and the others, which breaks the original SU(3) symmetry down to SU(2)$\times$U(1).
%Systems symmetric under the special unitary group of degree $\mathcal{N}$ higher than 2 are attracting growing interest in recent years due to the theoretical prediction of exotic ground states and the experimental realization of ideal SU($\mathcal{N}$)-symmetric interactions with the use of cold alkaline-earth(-like) atoms. Whereas the ground state has been extensively studied for different values of $\mathcal{N}$ and different lattice geometries in previous theoretical works, the effects of thermal fluctuations and ``external fields'' that partially break the SU($\mathcal{N}$) symmetry have been rarely investigated~\cite{Yamamotosan}, in contrast to the SU(2) spin systems under external magnetic fields. In cold-atom experiments, such an external-field effect can be simulated by imposing a global imbalance of populations among the $\mathcal{N}$ components~\cite{spinimbalance}. Here, we consider the case of $\mathcal{N}=3$ with population imbalance between one component and the others, which breaks the original SU(3) symmetry down to SU(2)$\times$U(1).
The system is described by the antiferromagnetic SU(3) Heisenberg model with an external field:
%------------------------
\begin{equation} 
\label{eq:Eq_the_SU(3)_Heisenberg_model}
\hat{\mathcal{H}}
=\frac{J}{2}\sum_{\langle i,j\rangle}\hat{{\bm{\lambda}}}_i \cdot \hat{\bm{\lambda}}_j  -D\sum_i \hat{\lambda}_{8,i} \ \ \ (J>0),   
\end{equation} 
%------------------------
%\noindent
where $\hat{\bm{\lambda}}_i=(\hat{\lambda}_{1,i}, \hat{\lambda}_{2,i}, \cdots,   \hat{\lambda}_{8,i})$ are the generators of the SU(3) Lie algebra in the defining representation~\cite{Gell_Mann_matrices}, acting on the three local basis states, which we refer to as three ``colors'', $|{\rm R}_i \rangle, |{\rm B}_i \rangle$, and $|{\rm G}_i \rangle$~\cite{SU(3)_1,SU(3)_2}, at site $i$. 
%of the square lattice. 
In the form given by the Gell-Mann matrices, $\hat{\lambda}_3$ and $\hat{\lambda}_8$ are diagonal, ${\rm diag}[1,0,-1]$ and $\frac{1}{\sqrt{3}} {\rm diag} [1,1,-2] $, respectively, while the others have off-diagonal entries, responsible for the color change of the local state. The last term of Eq.~(\ref{eq:Eq_the_SU(3)_Heisenberg_model}) represents a bias field controlling global population imbalance $P_{\rm g}$ between $\{{\rm R},{\rm B}\}$ and ${\rm G}$. Although this partially breaks the original SU(3),
%is partially broken by the external field $D$, 
the system still possesses the continuous SU(2)$\times$U(1) symmetry, related to the global rotations in the SU(2) space generated by $\hat{\lambda}_{1-3}$ and that around $\hat{\lambda}_{8}$.         

% %------------------------
% %表
% \begin{table}[hbtp]
%  \label{Variational_parameters}
%   \caption{Variational parameters of the five states in Fig.~\ref{fig:1} for $D \neq 0$ up to SU(2)$\times$U(1) transformations. The  wave vectors, ${\bm{Q}}_2$ and ${\bm{Q}}_3^{\pm}$, denote $(\pi, \pi)$ and $(2\pi/3, \pm 2\pi/3)$, respectively.}
%   \label{table:1}
%   \centering
  
%   \begin{tabular}{lcccc}
%     \hline 
%     \hline 
%     state 
%     &  
%     &$r_i=r$  
%     & $\{\theta_i,\phi_i\} $ 
%     & $\alpha_i$ \\
%     \hline
%     (i)     
%     &
%     & $0$ 
%     & --- 
%     & --- \\ \\
    
%     (ii)    
%     &
%     & $\frac{1}{2}(1+\frac{\sqrt{3}}{8}\frac{D}{J})$ 
%     & $A$
%     & ${\bm{Q}}_2\cdot{\bm{r}}_i$ \\ \\
    
%     (iii)   
%     &$(D<0)$ 
%     & $\frac{1}{2}( 1+\frac{1}{\sqrt{3}}\frac{D}{J} ) $ 
%     & $B^{\pm}$
%     & ${\bm{Q}}_3^{\mp} \cdot {\bm{r}}_i$ \\ 
    
%     &$(D>0)$ 
%     & $\frac{1}{2}( 1-\frac{1}{2\sqrt{3}}\frac{D}{J} ) $
%     & $B^{\pm}$
%     & ${\bm{Q}}_3^{\mp} \cdot {\bm{r}}_i$ \\ \\
    
%     (iv)   
%     & 
%     & $1$  
%     & $C$
%     & --- \\ \\
    
%     (v) 
%     &
%     & 1 
%     & $B^{\pm}$
%     & --- \\
%     \hline
%     \hline 
%   \end{tabular}
% \end{table} 
% % --------------------
%------------------------
% 画像
\begin{figure}[t]
\includegraphics[width=6.65cm]{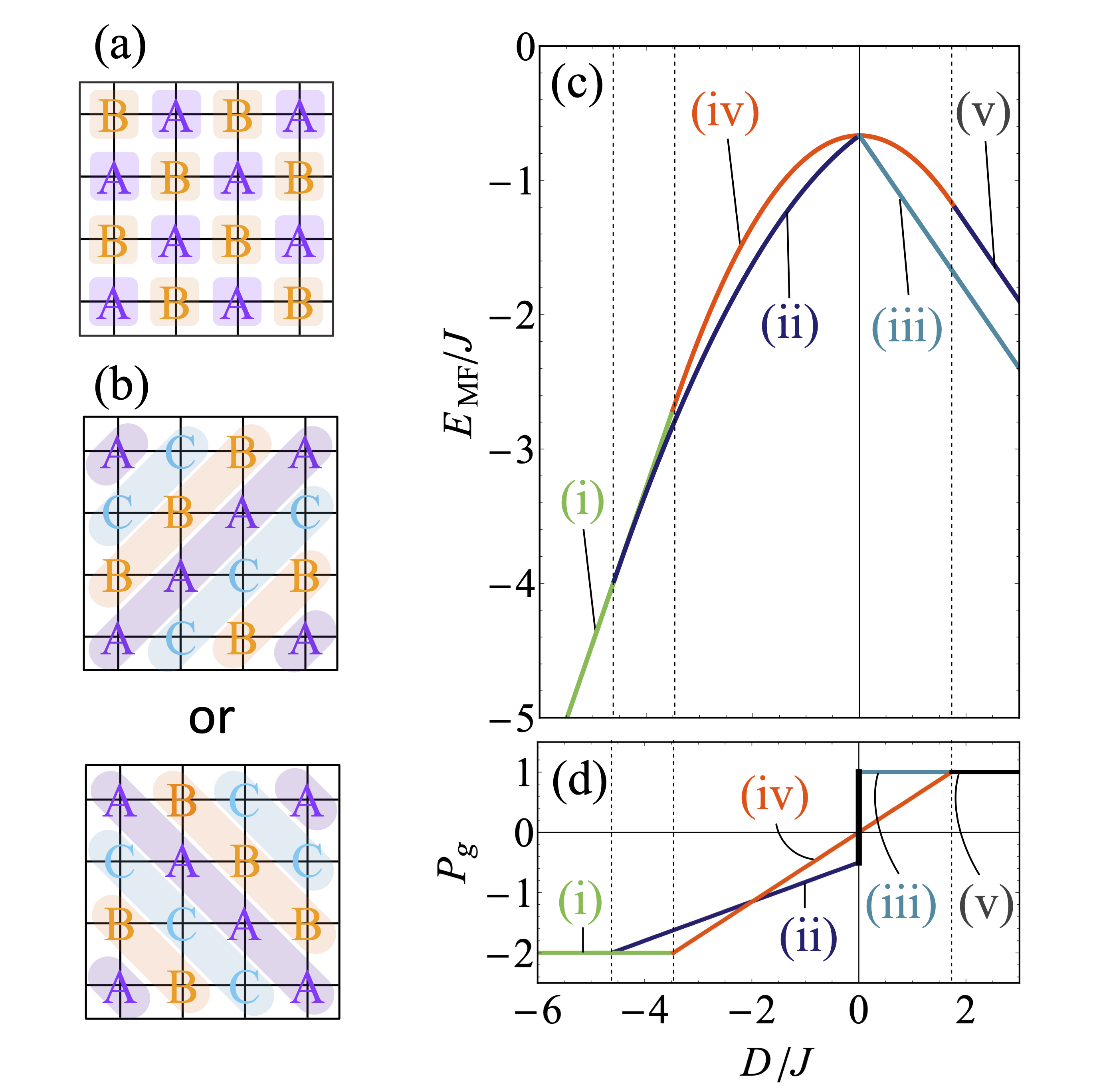}
\caption{\label{fig:1}
Illustration of the two and three-sublattice structures with ordering wave vectors (a) $\bm{Q}_2=(\pi, \pi)$ and (b) $\bm{Q}_3^{-}=(2\pi/3,-2\pi/3)$ or $\bm{Q}_3^{+}=(2\pi/3,2\pi/3)$. {Sublattices are labeled A, B, C.} (c) Energy per site and (d) global population imbalance of the mean-field states as functions of the field $D$ at  zero temperature.
}
\end{figure}
%------------------------
%------------------------
% 画像
\begin{figure}[t]
\includegraphics[width=7cm]{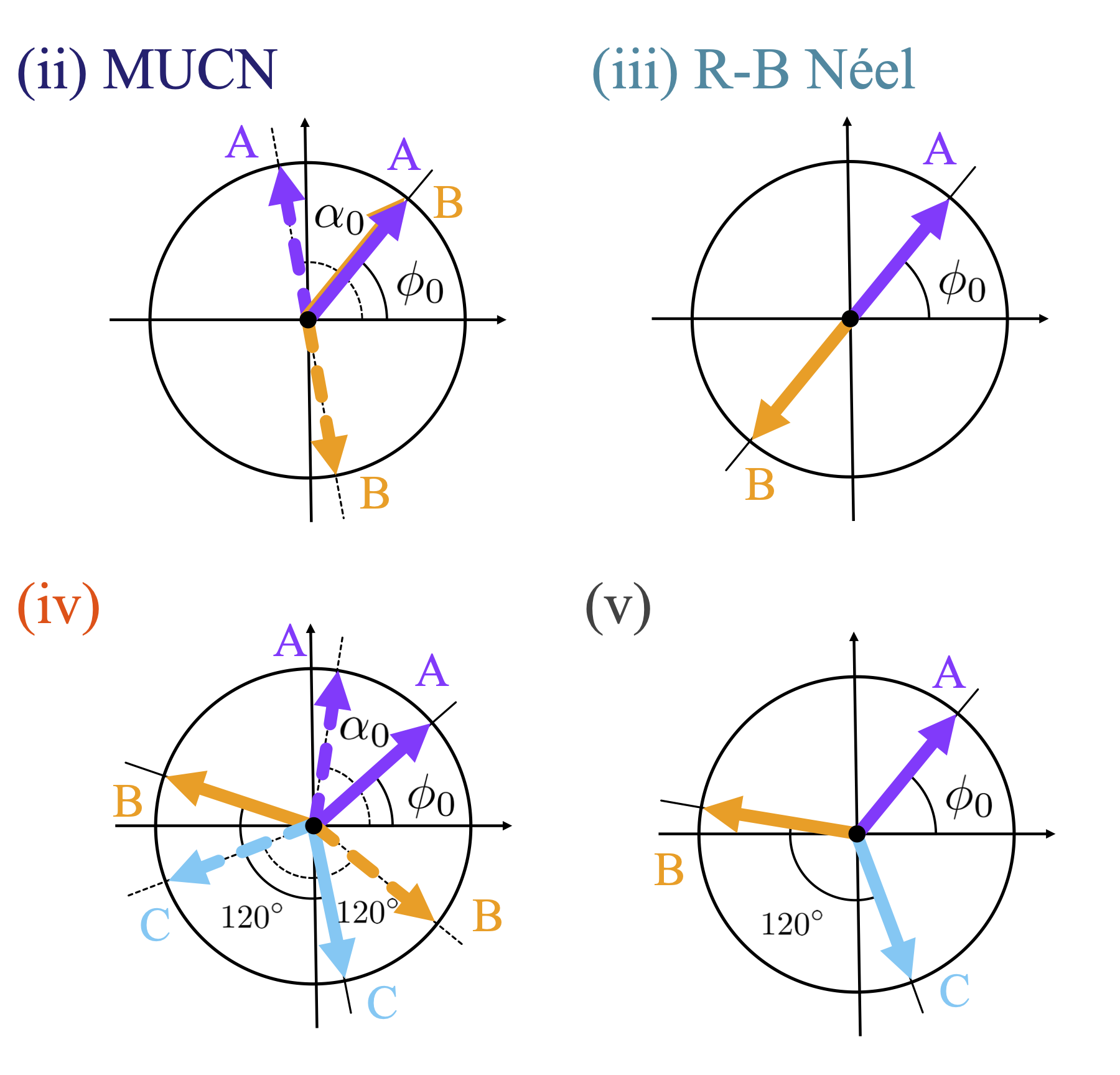}
\caption{\label{fig:2}
Two relative phases $\alpha_i,\phi_i$ in the mean-field states (ii)-(iv) on the corresponding sublattice structures displayed in Figs.~\ref{fig:1}(a) or \ref{fig:1}(b) in a fixed gauge with $\theta_i=\pi/2$. The angles $\alpha_0,\phi_0$ can be arbitrarily chosen independently. In states (iii) and (v), $\alpha_i$ is not defined since $s_i=1$.
}
\end{figure}
%------------------------

%\textit{Experimental realization.---} 
The SU(3) Heisenberg model corresponds to the spin-1 Hamiltonian with equal bilinear and biquadratic exchange couplings~\cite{BBQ_1}, which has been discussed in the context of spin liquid  in ${\rm NiGa}_2{\rm S}_4$~\cite{Spin_Nematic_Phase_in_S_1_Triangular_Antiferromagnets}, under the identifications $ \{ |{\rm R}_i\rangle, |{\rm B}_i \rangle, |{\rm G}_i \rangle \} \mapsto \{ |{1}_i \rangle, |{-1}_i \rangle, |{0} _i \rangle \}$. In the language of solid state physics, the imbalance field $D$ corresponds to the intrinsic single-ion anisotropy of the magnetic material. 
In artificial quantum systems of alkaline-earth(-like) atoms in optical lattices, the Hamiltonian (1) can be realized more directly with no fine-tuning of coupling parameters. Those atoms possess SU($2I+1$)-symmetric repulsive interaction for nuclear spin $I$ ($I=5/2$  for $^{173}{\rm Yb}$ and $I=9/2$ for $^{87}{\rm Sr}$), and the technique of optical pumping allows for the preparation of any number  $\mathcal{N}$  of components out of the $ 2I+1$ spin states.~\cite{taie-10}

% --------------------

\par
\textit{The mean-field ground state.}
First, we consider the mean-field ground state of Eq.~(\ref{eq:Eq_the_SU(3)_Heisenberg_model}) at zero temperature, based on a variational wave function of the form $ | \Psi \rangle = \prod_{i}{{| \psi}_i\rangle} $ with $|\psi_i\rangle
=d_{{\rm R},i} |{\rm R}_i\rangle+d_{{\rm B},i} |{\rm B}_i\rangle+d_{{\rm G},i} |{\rm G}_i\rangle$. The coefficients can be represented by a normalized complex vector parametrized as 
%------------------------
\begin{align}
    {\bm d}_i 
    &\equiv (d_{{\rm R},i} ,d_{{\rm B},i},d_{{\rm G},i}) \notag\\
    &=\left(\sqrt{s_i}\cos{\frac{\theta_i}{2}},\sqrt{s_i}{\rm e}^{i\phi_i}\sin{\frac{\theta_i}{2}},{\rm{e}}^{i\alpha_i}\sqrt{1-s_i}\right).
    \label{eq:2}
\end{align}
%------------------------
% with normalization $(|\bm{d}_i|^2=1)$. 
 The {amplitude and phases} $(s_i, \theta_i, \phi_i,\alpha_i) \in [0,1]\times [0,\pi] \times[0,2\pi)^2$ should be determined in such a way that the variational energy
%------------------------
$
E_{\rm MF}
=\langle\Psi|\hat{\mathcal{H}}|\Psi\rangle
=J\sum_{\langle i,j\rangle}|\bm{d}_{i}^{\dagger}\cdot \bm{d}_j|^2-D\sum_{i}(3s_i-2)/\sqrt{3}
$ 
%(** the last term should be $2r_i-2$ ? **)
%------------------------
is minimized. For $D=0$, namely at  the SU(3)-symmetric point, the minimization  only imposes that the $\bm{d}_i$ vectors on neighboring sites are orthogonal. This is not enough 
%on the square lattice 
to uniquely determine the global configuration of the $\bm{d}_i$ vectors, resulting in an accidental ground state degeneracy within the mean-field approximation. Previous studies have predicted that the thermal fluctuations favor the N\'eel configuration shown in Fig.~\ref{fig:1}(a) via entropic selection~\cite{SU(3)_1,SU(3)_2}, while the three-color stripe long-range order shown in Fig.~\ref{fig:1}(b) is chosen via a quantum order by disorder mechanism.

In the presence of an imbalance field $D$, this degeneracy is already lifted without taking fluctuations into account.
In Figs.~\ref{fig:1}(c) and \ref{fig:1}(d), we show the values of the energy per site and the global population imbalance $P_{\rm g}$, respectively, as functions of $D/J$, for several mean-field solutions, labeled (i)-(v).
For $D<0$, the imbalance field $D$ tends to increase the global population of the $|{\rm G}\rangle$ state and competes with the antiferromagnetic coupling $J$, which tends to arrange different colors at neighboring sites.
%; this is expected to lead to exotic orders and phase transition.
When $D$ is sufficiently negative ($D<-8J/\sqrt{3}$), the forced ferromagnetic state [(i) in Fig.~\ref{fig:1}] is formed ($|\psi_i\rangle=|{\rm G}\rangle$ for all $i$). For $-8J/\sqrt{3}<D<0$, we found  the ``minority-united canted-N\'eel'' (MUCN) phase [(ii) in Figs.~\ref{fig:1} and \ref{fig:2}], in which the variational parameters in Eq.~(\ref{eq:2}) are $(s_i, \theta_i, \phi_i, \alpha_i)=(\frac{1}{2}\left(1+\frac{\sqrt{3}}{8}\frac{D}{J}\right),\theta_0,\phi_0,\bm{Q}_2\cdot\bm{r}_i+\alpha_0)$, as the ground state. Here, {the phases} $\theta_0,\phi_0$, and $\alpha_0$ are arbitrary and $\bm{Q}_2=(\pi,\pi)$ {is the propagation vector}. In the MUCN state, while the phase $\alpha_i$ is alternating on the two sublattices (see Fig. \ref{fig:1}(a)), interestingly the SU(2) sector of R and B exhibits ferromagnetic order with uniform $\theta_0$ and $\phi_0$, despite the antiferromagnetic nature of the exchange interaction. This can be understood as a mechanism to minimize the interaction energy between the majority G and the minorities $\{ {\rm R},{\rm B} \}$. Notice that, the MUCN state can be mapped to the conventional canted-N\'eel state 
of quantum antiferromagnets under magnetic field through the mapping 
%via 
$\cos{(\theta/2)}|{\rm R}\rangle+{\rm{e}}^{i\phi}\sin{(\theta/2)}|{\rm B}\rangle \mapsto |\uparrow \rangle$, $ |{\rm G}\rangle \mapsto|\downarrow \rangle$.
For $D>0$, since the global population of $ \{ |{\rm R}_i\rangle, |{\rm B}_i \rangle \}$ tends to  increase, there is no competition with the antiferromagnetic interactions. Therefore, an infinitesimally small field stabilizes (iii), the standard N\'eel order (with R and B), which is of course the ground state of the SU(2) Heisenberg model. 

Although the mean-field theory predicts the two-sublattice ground states (MUCN and R-B N\'eel), besides the uniform state for $D<-8J/\sqrt{3}$, we also found three-sublattice solutions as metastable states [corresponding to (iv) and (v) in Figs.~\ref{fig:1}(c) and \ref{fig:1}(d)]. While in state (iv) the three colors coexist with $r(D<0)=\frac{1}{2}\left(1+\frac{1}{\sqrt{3}}\frac{D}{J}\right)$, $r(D>0)=\frac{1}{2}\left(1-\frac{1}{2\sqrt{3}}\frac{D}{J}\right)$, the component $|\rm G\rangle$ vanishes in state (v), that is, $s_i=1$. All these states, including the above-mentioned MUCN and R-B N\'eel, are massively degenerate, owing to the SU(2)$\times$U(1) symmetry of the Hamiltonian. Upon choosing the gauge $\theta_i=\pi/2$, for which densities of B and R particles are identical and uniform on the lattice, we summarize the remaining two parameters $\phi$ and $\alpha$ of the mean-field solutions [(ii)-(v)] in Fig.~\ref{fig:2}. 
%As mentioned earlier, the previous studies predicted that the ground state at $D=0$ develops three-color long-range order as a result of long-wavelength quantum fluctuations. 
%in the vicinity of $D=0$. 
%We focused on two-sublattice state because they are induced within a finite temperature analysis. [**??**]

%けっこう複雑だけど、theta=pi/2を含むので、そのgaugeでのalphaとphiを書くin Fig2、ちなみに、2subも同じ感じで描いた
%3subはquantumfkucでもでるかもだけど、ここではthermalを考えるので
%どの状態もmaniholdが大きい
%~captionで言っているkとをいう。

%\textit{Semi-classical Monte-Carlo simulations.---}
\textit{Thermal phase diagram.}
{Since the lowest temperatures reached in  current experiments are in the range $T\simeq 0.7 J$ [SU(6) chain] \cite{taie-22} to $T\simeq 0.9 J$ [SU(2) square lattice] \cite{Short_range_antiferro}}, addressing the finite temperature effects becomes important beyond a purely theoretical motivation.
%It is important to address finite temperatures $T>0$ because the current experiments have still not reached the highly-quantum low temperature region. 
We hereby take into account thermal fluctuations from the mean-field ground state  using  semi-classical multi-color Monte Carlo (sMC) simulations~\cite{sMC, Yamamotosan,remund-22}. 

We perform the standard Metropolis algorithm that allows the complex vectors $\bm{d}_i$ on $L \times L$  sites under periodic boundary conditions to fluctuate thermally with the Boltzmann weight $\exp{(-E_{\rm MF}/T)}$. In addition, we employ the relaxation acceleration (RA) technique~\cite{Yamamotosan}
with local unitary transformations $ \exp{( ic \hat{ \mathcal{H}}_{i}^{\rm loc} / \| \hat{\mathcal{H}}_{i}^{\rm loc}} \|_{\rm F}) $ defined by the local mean-field Hamiltonian $\hat{\mathcal{H}}_{i}^{\rm loc} =\frac{J}{2}\sum_{j \in {\rm{NN}}_i} \langle \hat{\bm{\lambda}}_j \rangle \cdot \hat{\bm{\lambda}}_i -D\hat{\lambda}_{8,i}$. Here the real number $c$ is randomly generated from an uniform distribution on the interval $[-1,1]$, $ \|\cdots\|_{\rm F} $ is the Frobenius norm, and the sum 
%$\sum_{j \in {\rm NN}_i}$ 
runs over all nearest-neighbor sites of site $i$. In the case of the SU(3) system, the eigenvalues of $ \hat{\mathcal{H}}_{i}^{\rm loc} $ can be calculated analytically, which contributes to the reduction of the calculation cost~\cite{diagonalization}.
The single Monte Carlo step consists of the Metropolis updates for $d_{\sigma,i}$ over the entire lattice, followed by  double RA sweeps. This procedure can enhance decorrelation and increase the convergence speed~\cite{Yamamotosan}.

\begin{figure}[tb]
\includegraphics[width=8.5cm]{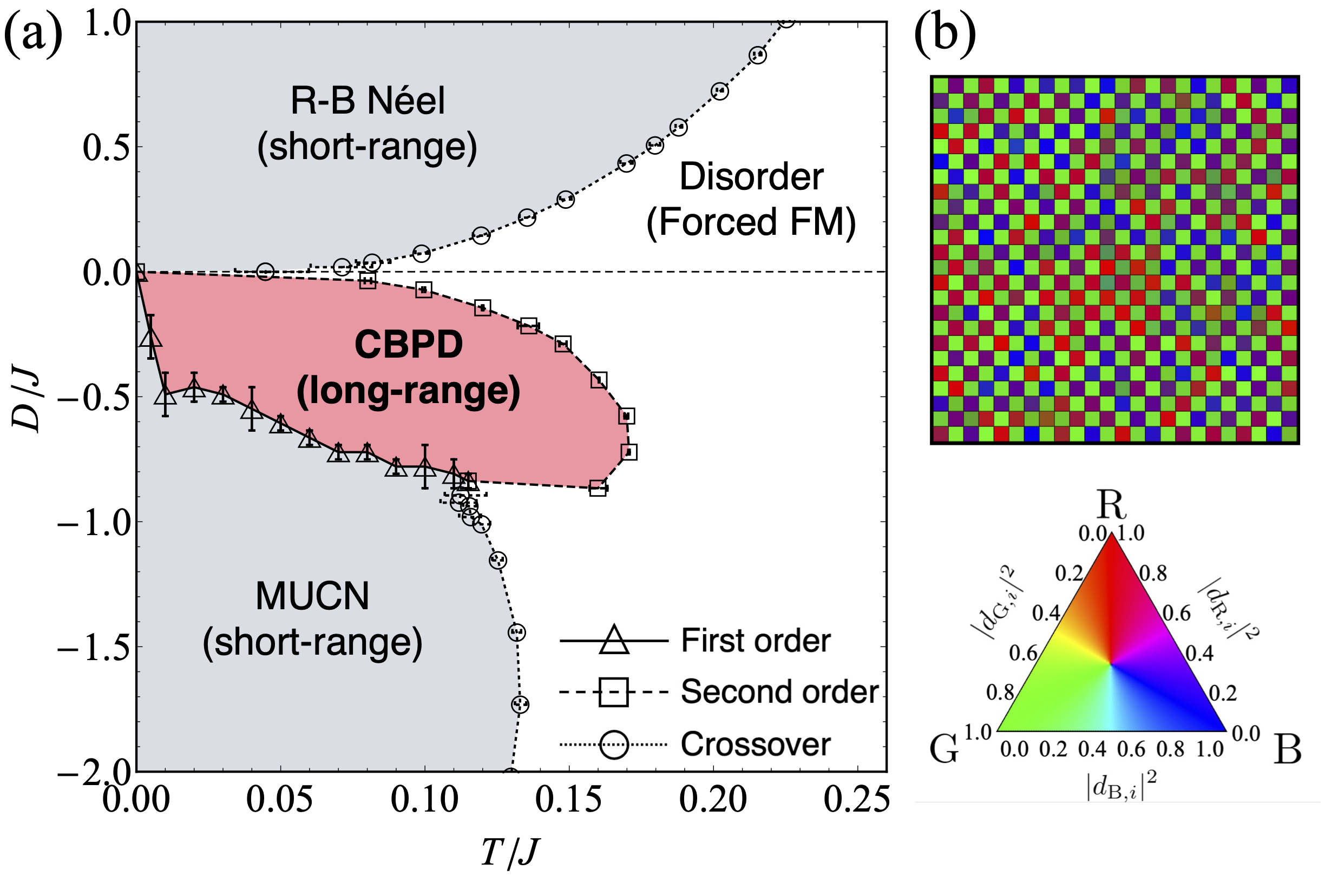}
\caption{\label{fig:3}
(a) Thermal phase diagram obtained by the sMC simulations. The crossover temperature $T_{\rm c}^{\ast}$ is determined by the temperature where the correlation length reaches ten sites. (b) A snapshot of the checkerboard partially disordered state (CBPD) order from the Monte Carlo simulations.
}
\end{figure}

%\textit{Thermal phase diagram.---}
Figure~\ref{fig:3} shows the sMC thermal phase diagram. The most surprising feature is a fluctuation-induced phase with true long-range order emerging for $-0.8 \lesssim D <0$. The R-B N\'eel and MUCN states, being continuous-symmetry-breaking states, only possess short-range correlations and a crossover connects them to the high-temperature disordered phase.
%The mean-field  states [(ii), (iv)] break the continuous symmetry with respect to global spin rotations, which forbids finite temperature phase transitions according to the Mermin-Wagner-Hohenberg theorem. Therefore the correlation that the ground state (R-B Néel, MUCN) possesses remains only in short range for $T>0$. 
%Let us note that, for the nature of sMC, this phase diagram is most reliable away from the highly-quantum (low-temperature) regime.
%which is basically reliable in the region away from the highly-quantum regime.
%-------------------------------%
%画像
%-------------------------------%

\begin{figure}[]tb]
\includegraphics[width=7cm]{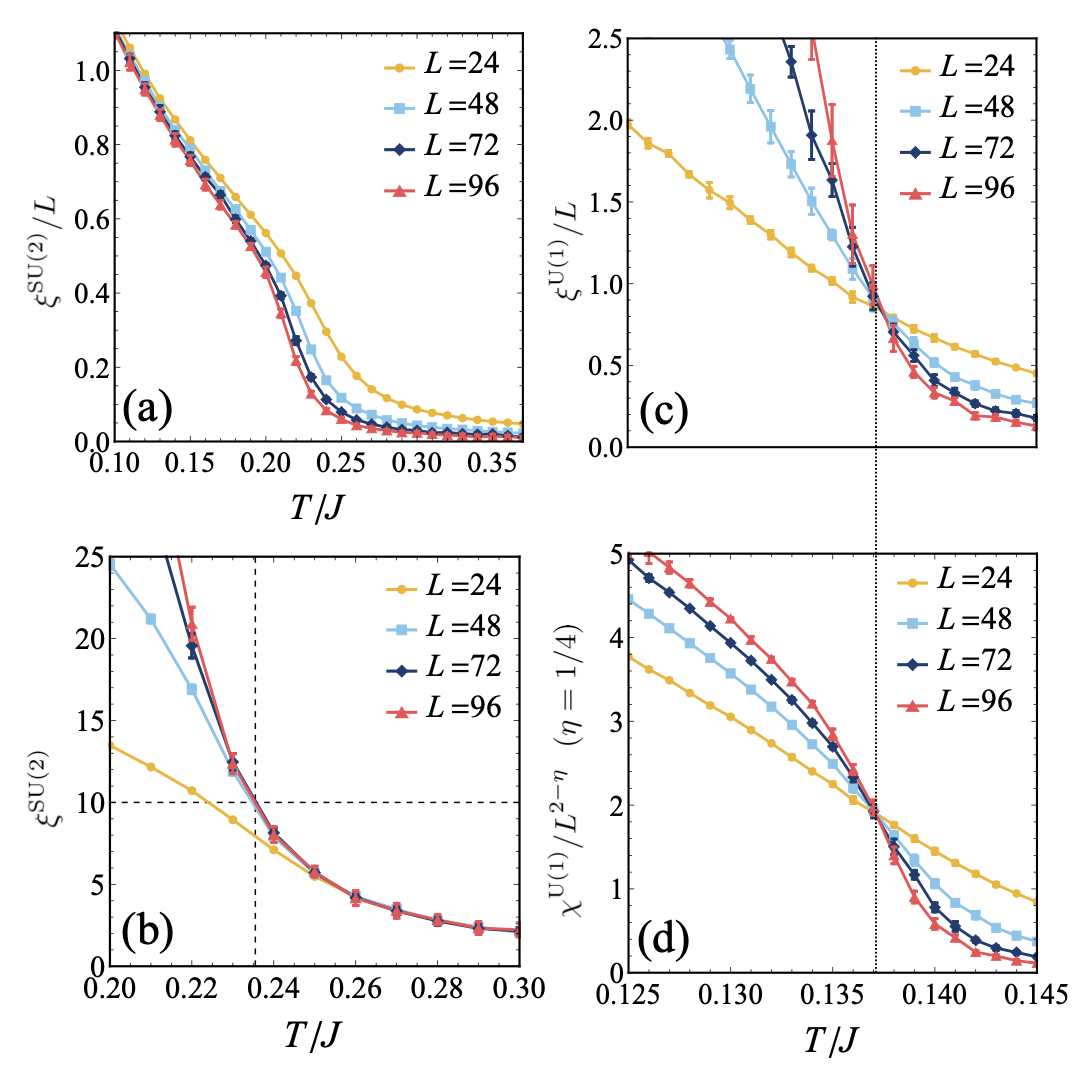}
\caption{\label{fig:4}
Typical examples of the numerical data for the scaling analyses that obtain the finite-temperature phase diagram shown in Fig.~\ref{fig:3}. (a) scaled correlation length $\xi^{\rm SU(2)}/L$ and (b) correlation length $\xi^{\rm SU(2)}$ along $D/J=2/\sqrt{3}$. (c) scaled correlation length and (b) scaled spin susceptibility with respect to $\hat{\lambda}_8$ at $D/J=-1/2\sqrt{3}$. 
}
\end{figure}

In order to determine the transition (or crossover) lines in Fig.~\ref{fig:4}, we calculate the correlation lengths~\cite{Seabra,Yamamotosan}
%-------------------------------%
$
 \xi(\bm{q})=|\delta \bm{q}|^{-1}
 \sqrt{[\mathcal{S}(\bm{q})/\mathcal{S}(\bm{q}+\delta \bm{q})]-1},
$
%-------------------------------%
where $\mathcal{S}(\bm{q})$ is the structure factor for wave vector $\bm{q}$ and $\delta\bm{q}=(2\pi/L,0)$. 
%[** notation: maybe $\delta\bm{q}? $**]
We use 
%-------------------------------%
$
 \mathcal{S}^{\rm SU(2)}(\bm{q})=\sum_{\mu=1}^{3}\left\langle\left|\sum_{i}\hat{\lambda}_{\mu,i}{\rm{e}}^{i\bm{q}\cdot \bm{r}_i}\right|^2/L^2\right\rangle
$
%-------------------------------%
and 
%-------------------------------%
$
 \mathcal{S}^{\rm U(1)}(\bm{q})=\left\langle\left|\sum_{i}\hat{\lambda}_{8,i}{\rm{e}}^{i\bm{q}\cdot \bm{r}_i}\right|^2/L^2\right\rangle
$
%-------------------------------%
to detect the correlations in the SU(2) and U(1) sectors, respectively. 
The R-B N\'eel and MUCN phases can be identified by $\mathcal{S}^{\rm SU(2)}(\bm{q})$ with  ordering vectors $\bm{q}=\bm{Q}_2 \equiv (\pi,\pi)$ and $\bm{q}=\bm{0}$, respectively. We confirm that there is no crossing point of $\xi^{\rm SU(2)}/L$ for different system sizes down to low temperatures [see a typical case in Fig.~\ref{fig:4}(a)], which indicates no long-range order at $T>0$ as expected. 
%-------------------------------%
%画像
%-------------------------------%
%
In Fig.~\ref{fig:3}, we plot the crossover temperature $T_{\rm c}^{\ast}$ from disordered to short-range correlated state by choosing the condition $\xi^{\rm SU(2)}=10$ sites [see Fig.~\ref{fig:4}(b)], which is comparable to the linear system size in typical cold-atom experiments in two dimensions~\cite{Short_range_antiferro}.

The emergence of the nontrivial true long-range order  between the regions with the R-B N\'eel and MUCN short-range correlations for $T>0$ is of particular significance.
%Of particular interest is that the nontrivial true long-range order that was absent in the mean-field ground state emerges in between the regions with the R-B Néel and MUCN short-range correlations for $T>0$.
In this phase, most of the snapshots in the sMC simulations show a checkerboard pattern where one sublattice is occupied mostly by $|{\rm G}\rangle$ and the other by random superpositions of $|{\rm R}\rangle$ and $|{\rm B}\rangle$: $|\psi_{i_{\rm A}}\rangle \simeq |{\rm G}\rangle,|\psi_{i_{\rm B}}\rangle \simeq \cos{(\theta_{i_{\rm B}}/2)}|{\rm R}_{i_{\rm B}}\rangle + {\rm{e}}^{i\phi_{i_{\rm B}}}\sin{(\theta_{i_{\rm B}}/2)}|{\rm B}_{i_{\rm B}}\rangle$, where $\theta_{i_{\rm B}},\phi_{i_{\rm B}}$ are randomly chosen 
%from the Bloch sphere 
at each site in each sMC snapshot [i.e., disordered in the SU(2) sector]. This state breaks neither  SU(2) nor U(1) symmetry, but  {\it only the discrete translational symmetry}. Thus, the Mermin-Wagner-Hohenberg theorem~\cite{MerminWagner,Hohenberg} does not forbid a finite-temperature phase transition to this checkerboard partially disordered state' (CBPD) with true long-range order. 
In fact, as seen in Fig.~\ref{fig:4}(c), the scaled correlation lengths $\xi^{\rm U(1)}/L$, obtained from $\mathcal{S}^{\rm U(1)}(\bm{q}=\bm{Q}_2)$, for different linear sizes $L$ cross each other at a certain critical point $T_{\rm c}>0$. According to the finite size scaling theory the scaled magnetic susceptibility $\chi^{\rm U(1)}/L^{2-\eta}$, where  $\chi^{\rm U(1)}=(J/k_{\rm B}T)\mathcal{S}^{\rm U(1)}(\bm{Q}_2)$, should become size-independent at the critical temperature $T_{\rm c}$: Fig.~\ref{fig:4}(d) shows that   this scaling law is reproduced with critical exponent $\eta=1/4$,  indicating the 2D Ising universality class, as expected from the $Z_2$ translational symmetry breaking of the checkerboard pattern.

Let us discuss the mechanism of the emergence of CBPD order. Within the mean-field approximation, this phase is included in the infinitely degenerate ground-state manifold at the SU(3)-symmetric point, since each pair of the $\bm{d}_i$ vectors on neighboring sites is orthogonal. For $T>0$ one sublattice is  disordered in the SU(2) sector while the checkerboard pattern is kept. Since the partial disorder compensates the loss of entropy, thermal fluctuations select this phase from the zero-temperature degenerate manifold.~\footnote{We conjecture that this argument should be true also for $\mathcal{N}>3$, that is the SU($\mathcal{N}$) Heisenberg model with $-D\sum_i \hat{\lambda}_{\mathcal{N}^2-1,i}$ field term. } {Let us also recall that the three-sublattice state selected by purely quantum fluctuations is unstable for any $T>0$, according to linear flavor-wave theory \cite{SU(3)_1}, in favor of two-sublattice states, and therefore the CBPD phase is expected to be robust against the inclusion of quantum effects beyond single site.
in the three-dimensional case, the entropic selection of two-sublattice phases has also been reported \cite{sotnikov-14}, 
%although the MWH no-go rule is not applied. 
although the three-sublattice phase is not completely excluded near zero temperature due to the higher dimensionality.}

\begin{figure}[tb]
\includegraphics[width=6cm]{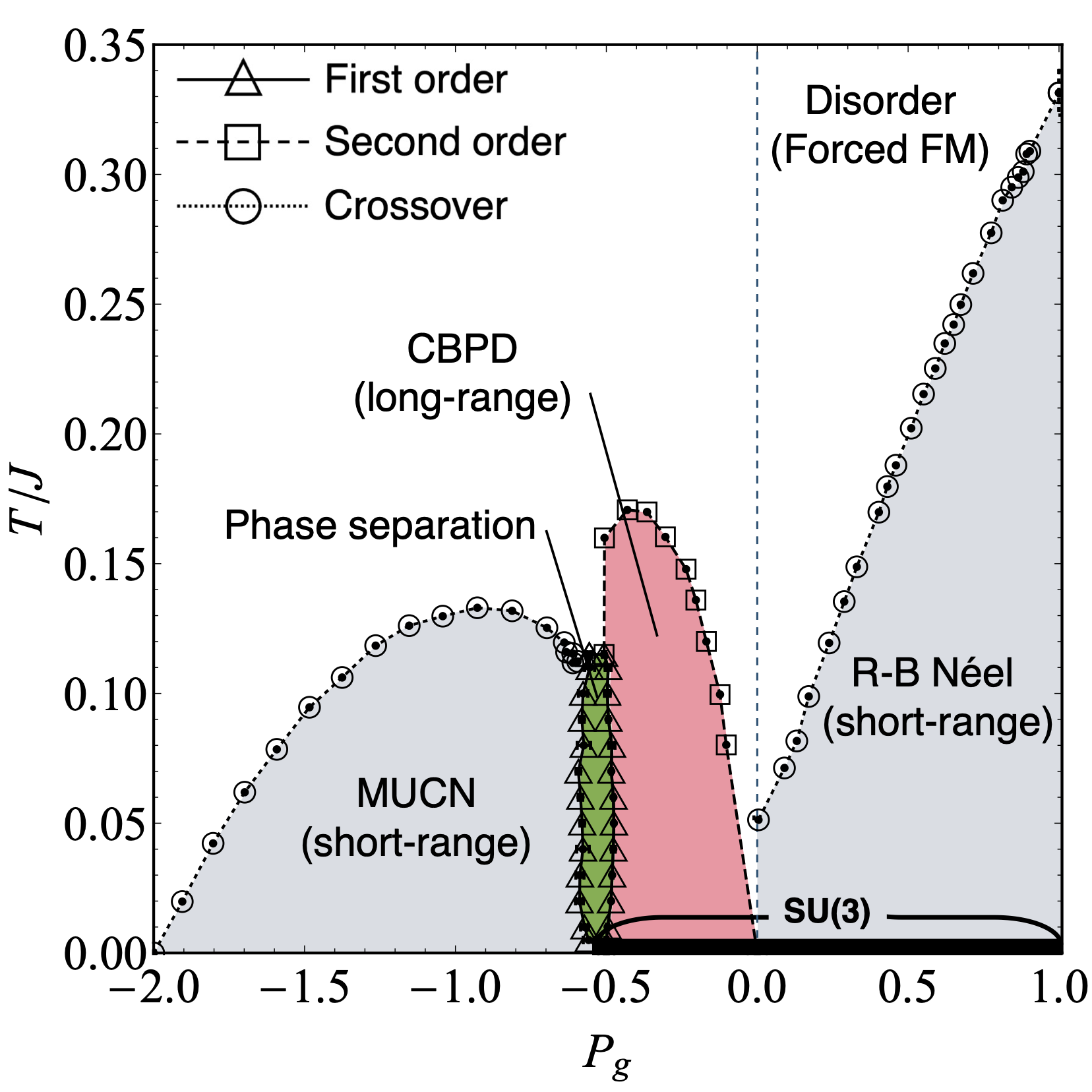}
\caption{
\label{fig:5}
Thermal phase diagram as a function of the population imbalance $P_{\rm g}$. The thick horizontal line corresponds to the SU(3)-symmetric point. Along that line, the variational solution is highly degenerate.
}
\end{figure}

\textit{Experimental detection.} 
From an experimental perspective, it is convenient to describe the thermal phase diagram (Fig.~\ref{fig:3}) as a function of the global population imbalance $P_{\rm g}$ defined by $P_{\rm g} = \sqrt{3} \sum_i{\langle \hat{\lambda}_{8,i} \rangle} = \frac{ N_{\rm R}+N_{\rm B}-2N_{\rm G}}{N_{\rm R}+N_{\rm B}+N_{\rm G}} $, where $N_{\mu}$ represents the global population of color $\mu$~\cite{spinimbalance}, because this quantity, rather than $D$, is controllable  in cold atom experiments~\cite{spinimbalance}. Figure~\ref{fig:5} shows the $T$-$P_{\rm g}$ phase diagram. Note that a phase separation region appears between CBPD and MUCN.
%-------------------------------%
% 画像
%-------------------------------%
The detection of the CBPD order may be realized as follows. First, the $|\rm G \rangle$ state is removed from the entire lattice via  optical pumping \cite{taie-10} and  {horizontal pairs of neighboring  lattice sites are merged into single sites so as to form a new square lattice. We then look for signatures of the Mott insulating state, in which ideally all sites are singly occupied by either R or B state, with the established techniques~\cite{jordens-08,sherson-10,greif-11}. The experiment is repeated and the same protocol is  operated with vertical pairs instead. If, in both instances, the Mott insulating state is detected, this would be proof of the realization of the CBPD order.}
%and perturbative lattice modulation~\cite{Single_atom_resolved_fluorescence_imaging_of_an_atomic_Mott_insulator,Probing_Nearest_Neighbor_Correlations_of_Ultracold_Fermions_in_an_Optical_Lattice}. The $|\rm G \rangle$ state is optically removed from the entire lattice; then neighboring pairs of lattice sites are merged into single sites, followed by perturbative lattice modulation. As a result, the CBPD order should show clear signatures of the formation of the Mott-insulating phase.[**?**] 
In addition, the eventual extension of the quantum gas microscope technique to SU($\mathcal{N}$) systems~\cite{okuno-20} could detect each phase more directly.
Note that the sMC method replaces the statistics of discrete quantum levels by approximate continuous classical statistics of the vectors $\bm{d}_i$. Since this treatment should overestimate the entropy of paramagnetic states, thus underestimating  the transition/crossover temperatures, the lines of $T_{\rm c}(T_{\rm c}^{\ast})$ in Fig.~\ref{fig:5} should be taken as  lower bounds. 

\textit{Conclusion.}
We have investigated the emergence of long-range order by thermal disorder in the square-lattice SU(3) Fermi gas at strong coupling with population imbalance and proposed experimental setups to realize and detect our predictions with alkaline-earth(-like) atoms. Applying mean-field theory and the sMC method to the SU(3) Heisenberg model, we found that the competition between antiferromagnetic interactions and the population imbalance gives rise to several interesting magnetic states, including a true long-range ordered state that breaks only a discrete symmetry.
%despite the high continuous symmetry of the system. 
This state enjoys an entropic advantage coming from the abundance of color degrees of freedom and is therefore favored by thermal fluctuations. This scenario generalizes the idea of emergent discrete symmetry breaking in low-dimensional systems without the traditional forms of magnetic frustration. Furthermore, we recall that the quantum fluctuations at $D=0$ (no population imbalance) select the three-color stripe order, which should then persist for sufficiently small $D<0$. This creates the condition for a finite-temperature transition originating from the difference between quantum and thermal order-by disorder selection, a very rare phenomenon in magnetic systems~\cite{Sheng-92,yamamoto-19}.
%it would be worth investigating this aspect further, both experimentally and with theoretical methods that deal with quantum and thermal fluctuations simultaneously.

In conclusion, our work offers new directions in SU($\mathcal{N}>2$) magnetism, which is presently under intensive investigation in cold-atom experiments, showing further examples of the rich physics induced by population imbalance. It also provides a new insight into the phase transition phenomena of highly-symmetric systems, which can prove beneficial in other contexts, such as  solid-state physics~\cite{Spin_Nematic_Phase_in_S_1_Triangular_Antiferromagnets}.  
%[** ``astrophysics, etc.'' ?**]

\begin{acknowledgments}
We would like to thank, Y. Takahashi, Y. Takasu and  S. Taie, for useful discussions. This work was supported by JSPS KAKENHI Grant No.~18K03525 (D.Y.), No. 21H05185 (G.M., D.Y.), No. 22H01171 (N.F., D.Y.), and JST PRESTO Grant No.~JPMJPR2118, Japan (D.Y.)
\end{acknowledgments}

\providecommand{\noopsort}[1]{}\providecommand{\singleletter}[1]{#1}%

\end{document}